# Discovering author impact: A PageRank perspective

**Erjia Yan[1], Ying Ding**

*School of Library and Information Science, Indiana University, 1320 East 10th Street, Bloomington, IN 47405-3907, United States. E-mail: {eyan, dingying}@indiana.edu*
*Tel: 1-812-606-8091*

**Abstract**

This article provides an alternative perspective for measuring author impact by applying PageRank algorithm to a coauthorship network. A weighted PageRank algorithm considering citation and coauthorship network topology is proposed. We test this algorithm under different damping factors by evaluating author impact in the informetrics research community. In addition, we also compare this weighted PageRank with the *h*-index, citation, and program committee (PC) membership of the International Society for Scientometrics and Informetrics (ISSI) conferences. Findings show that this weighted PageRank algorithm provides reliable results in measuring author impact.

---

[1] Corresponding author

# 1 Introduction

In bibliometrics, the number of citations is an indicator used to measure the impact of scientific publications. Authors whose publications have been intensively cited usually have a higher academic impact in their respective fields; however, there are situations where citations do not provide a full perspective on the impact of an author.

Coauthorship network analysis, with its sound theory and methodology derived from physics, mathematics, graph theory, and social sciences, is expected to serve as the complement to traditional citation analysis. Specifically, the micro-level metrics for coauthorship network analysis can inform us about the power, stratification, ranking, and inequality in social structures (Wasserman & Faust, 1994). Such an approach captures the features of the individual actors in a network along with consideration of the topology of the network.

Among many indicators, PageRank has great potential for coauthorship network analysis. The PageRank algorithm (Brin & Page, 1998) assumes web hyperlinks to be trust votes and ranks the search results based on these links interconnecting them. PageRank brings a new method to information retrieval for a better ranking of the web. For coauthorship networks, the PageRank algorithm gives higher weights to the authors who collaborate with different authors, and also to authors who collaborate with a few highly coauthored authors. PageRank is thus chosen as a complementary method to citation analysis, enabling us to identify author impact from a new perspective.

The field of informetrics was selected for this study, since it is a fast-developing discipline (Bar-Ilan, 2008). More importantly, it is a coherent field, in that the selection will not result in too many breakages of collaboration ties, which is of vital importance for the robustness of coauthorship networks. In the first section, related studies on the PageRank algorithm for bibliometrics are introduced, and the second section, PageRank algorithm as well as the weighted PageRank algorithm is presented. In the result section, we (1) calculate the correlation coefficient between PageRank ranks and citation ranks for authors in the network, (2) discuss PageRank values under different damping factors from 0.15 to 0.85 with a 0.1 increment, (3) rank authors with the weighted PageRank algorithm, and (4) compare PageRank with the *h*-index, citation and PC members.

# 2 Related studies

Coauthorship networks can illustrate authors' social capital in terms of collaboration in the chosen discipline (Yan & Ding, 2009). For example, through centrality studies, authors with high betweenness centrality have more opportunities to broker the flow of information and have a higher social capital (Burt, 2002). Yan and Ding (2009) found that degree centrality, closeness centrality, and PageRank also measure authors' impacts on the field as well as their social capital. Disciplinarity can affect authors' collaboration patterns. Moody (2004) pointed out that collaborations in quantitative studies are very common, since "specialists are often added to research teams to do the analyses" (p. 218); by comparison, theoretical or historical specialties have a lower rate of coauthorship. For example, PhD students in the natural sciences usually work closely with their advisors and are thus more likely to co-publish work, while social science students tend to work more independently (Moody, 2004). Therefore, when comparing collaboration patterns, disciplinarity needs to be considered as well.

Currently, the PageRank algorithm has been applied to three types of scientific networks to assess research output: paper citation networks, journal citation networks, and coauthorship networks. Chen, Xie, Maslov, and Render (2007) applied the PageRank algorithm to assess the relative importance of all publications in the Physical Review family of journals. They found that PageRank values and citations for each publication were positively correlated. Ma, Guan, and Zhao (2008) applied PageRank to the evaluation of research influence of countries in the field of Biochemistry and Molecular Biology. They constructed a paper citation network with 236,517 papers from 261 seed journals categorized as Biochemistry and Molecular Biology in Science Citation Index, and calculated each paper's PageRank value with a damping factor of 0.5. They found that citation and PageRank were highly correlated, with the correlation coefficient reaching to 0.9 at the 0.01 level.

Using journal data from the Institute for Scientific Information (ISI), Bollen, Rodriguez, and Van de Sompel (2006) demonstrated how a weighted version of the PageRank algorithm can be used to obtain a metric that reflects prestige. They contrasted the rankings of journals according to ISI impact factor and weighted PageRank, discovering that both measures have overlaps and differences. Dellavalle et al. (2007) studied dermatology journals using the weighted PageRank algorithm, which assigned greater weight to citations originating from more frequently cited journals. They found that the weighted PageRank algorithm provided a more refined measure of journal status and changed relative dermatology journal rankings.

Liu, Bollen, Nelson, and Van de Sompel (2005) applied degree, closeness, betweenness centrality, and PageRank to coauthorship in the digital libraries research community. They also proposed AuthorRank, a weighted PageRank algorithm which considered collaboration intensity. They discovered that, compared to centrality measures, PageRank and AuthorRank have a more precise match with Joint Conference on Digital Libraries program committee members. Inspired by the aggregate function for the generation of author ranking based on publication ranking (Sidiropoulos & Manolopoulos, 2006), Fiala, Rousselot, and Ježek (2008) proposed a new version of PageRank which incorporated both citation and coauthorship graph property, and applied it to the DBLP digital library. Comparing the results with ranks of the winners of the ACM E. F. Codd Innovations Award, they found that their new version PageRank ranking has better recall of award winners than the standard PageRank. In another important article, Maslov and Render (2008) discussed some caveats of extending the PageRank algorithm to citation networks. They argued that, unlike hyperlinks, citations could not be updated after publication, which made aging effects much more important in citation networks than in the web. Meanwhile, they also considered that the habits of scientists looking for relevant scientific literature were different from web surfers, in that they usually search within shorter depth and cite older publications.

## 3 Methodology

*3.1 Data in the study*

The term "informetrics" was introduced by Blackert, Siegel, and Nacke in the 1970s and gained popularity by the organization of the international informetrics conference in 1987 (Egghe & Rousseau, 1990). The field of informetrics, actually, started in the first half of the twentieth century with works by Lotka, Bradford, and Zipf (Egghe, 2005).

Tague-Sutcliffe (1992) defines informetrics as "the study of the quantitative aspects of information in any form, not just records or bibliographies, and in any social group, not just scientists" (p. 1). Egghe (2005) uses informetrics as "the broad term comprising all the metrics studies related to information science, including bibliometrics (bibliographies, libraries, …), scientometrics (science policy, citation analysis, research evaluation, …), webometrics (metrics of the web, the Internet or other social networks such as citation or collaboration networks), …" (p. 1311).

In a recent article published in the *Journal of Informetrics*, Bar-Ilan (2008) conducted a detailed review of the status quo of informetrics in the 21st century. In our study, we adopted her data collection method and retrieved the following query from the Web of Science database (retrieval time: Jan 31st, 2009; time span: default all years): TS=(Informetric* OR bibliometric* OR webometric* OR scientometric* OR citation analy* OR cocitation analy* OR co-citation analy* OR link analy* OR hyperlink analy* OR self citation* OR self-citation* OR impact factor* OR science polic* OR research polic* OR S&T indicator* OR citation map* OR citation visuali* OR information visual* OR h-index OR h index OR Hirsch index OR patent analy* OR Zipf OR Bradford OR Lotka OR collaboration network* OR coauthorship network* OR co-authorship network*) OR SO=(Scientometrics OR Journal of Informetrics). Most of these query words are limited to the subject category of INFORMATION SCIENCE & LIBRARY SCIENCE. After cleaning the irrelevant records, our data set finally included 5,096 papers[2] (articles and review articles) with 6,049 authors and 7,358 coauthor ties.

*3.2 PageRank for undirected graphs*

PageRank is a graph-based ranking algorithm used to determine the importance of a vertex within a graph by considering both its inbound links and outbound links (Ding et al., 2009). Although PageRank is originally designed for directed graphs, it can be applied to undirected graphs (Mihalcea, 2004; Perra & Fortunato, 2008). PageRank for undirected graphs has been used in computational linguistics, such as text summarization (Mihalcea, 2004), sentence extraction (Wang, Liu, & Wang, 2007), and word sense disambiguation (Mihalcea, Tarau, & Figa, 2004).

Adjacency matrix *A* is the basic matrix of a graph with the element $A_{ij}$ equal to 1 if node *i* and *j* are connected by a link and 0 otherwise. In coauthorship networks, nodes present authors, edges represent the coauthor relations, and the weights of the edges represent the coauthor frequency among these authors. The out-degree of a node is thus equal to its in-degree. For instance, if author *j* and author *k* are coauthored 3 times, then it is interpreted in the coauthorship graph as author *j* and author *k* having an edge with the weight of 3. In this case, our entry for the coauthorship matrix *A* is $A_{jk}=A_{kj}=3$. We set all the diagonal elements of *A* to zero. The PageRank algorithm will guarantee the matrix to be stochastic (each column sums to one) and irreducible (no non-zero entries).

---

[2]In the article titled "The effects of dangling nodes on citation networks" (2009 submitted), 99 records that have no cited references are deleted, which is resulted from the index problem of some records in the Web of Science database, and thus causing the difference.

*3.3 PageRank and weighted PageRank*

PageRank is not new to bibliometrics. Pinski & Narin (1976) proposed "influence weight", a metric utilizing eigenvalues of journal citation matrices, to measure the performance of journals or similar aggregates. PageRank is formally formulated by Page and Brin (1998), who developed a method for assigning a universal rank to web pages based on a weight-propagation algorithm. A page has high rank if the sum of the ranks of its inlinks is high (Brin & Page, 1998; Page, Brin, Motwani, & Winograd, 1999).

The PageRank of page *p* is given as:

$$PR(p) = \frac{(1-d)}{N} + d\sum_{i=1}^{k}\frac{PR(p_i)}{C(p_i)} \quad (1)$$

where *N* is the total number of pages on the web, $p_i$ is the page that links to *p*, *d* is damping factor, and $C(p_i)$ is the number of outlinks of $p_i$. PageRank of a page is conceived as the probability of a web surfer visiting the page after clicking on many links. The probability of a surfer which keeps clicking on links is thus given by the damping factor *d*. Since the surfer jumps to another page randomly after it stops clicking links, the probability therefore is implemented as the complementary part (1- *d*) into the algorithm (Ma, Guan, & Zhao, 2008).

The damping factor *d* in the original study by Brin and Page (1998) was set as 0.85. This value was prompted by the anecdotal observation that an individual will typically follow of the order of six hyperlinks, corresponding to a leakage probability (1- *d*) = 1/6≈0.15 (Chen, Xie, Maslov, & Render, 2007), "before becoming either bored or frustrated with this line of search and beginning a new search" (p. 9). This figure means that there is an 85% chance that a random surfer will follow the links provided by the present page, and a 15% chance that a random surfer starts a completely new page which has no link to previously surfed pages. In their empirical study (Chen et al., 2007), they also found that scientific papers usually follow a shorter path of about an average of two links, making the choice *d* = 0.5 more appropriate for citation networks. For retrieval purposes, it may not be appropriate to choose a small value for *d* because most of the pages will have similar probability. The topology of network will be dominant when choosing *d* close to 1; however, this will significantly increase the computing complexity (Boldi et al., 2005; Langville & Meyer, 2006). In this study, we calculate PageRank values under different damping factor 0.85, 0.75, 0.65, 0.55, 0.45, 0.35, 0.25, and 0.15 respectively, and verify to what extent varied damping factors will affect PageRank for coauthorship networks.

Weighted PageRank is not a new notion for scientific networks. Bollen et al. (2006) substituted the $C(p)^{-1}$ to the fraction of the journal's PageRank transferred to the journals it cites. Similarly, instead of $C(p)^{-1}$, Liu et al.'s (2005) version of weighted PageRank gives more weights to coauthor ties with fewer coauthors than those with large numbers of coauthors. Fiala et al.

(2008) proposed another weighted PageRank algorithm, and changed $C(p)^{-1}$ to $\sigma/\sum\sigma$, where $\sigma$ is a value between author *i* and author *j* and $\sum\sigma$ is the sum of $\sigma$ between author *i* and all authors. Embedded in $\sigma$ is another fraction which measures the numbers of citations from author *i* and author *j* and all citations from author *i* to the rest of the authors.

These weighted PageRank algorithms focus on the second part: $d\sum_{i=1}^{k}\frac{PR(p_i)}{C(p_i)}$, in which the topologies of networks are considered, yet the first part: $\frac{(1-d)}{N}$ is not discussed. The first part corresponds to equal probability of random surfing and is used to handle the RankSink problem (Haveliwala, 1999); this may be useful for a web crawler so as to cover extensive pages since the topology of the web cannot be obtained beforehand. But for coauthorship networks, this may not be helpful since the structures of networks are already captured. Hence, intuitively, influential authors should have a better chance to be randomly surfed. In this study, we incorporate citation counts with topology of network into the following formula:

$$PR\_W(p) = (1-d)\frac{CC(p)}{\sum_{j=1}^{N}CC(p_j)} + d\sum_{i=1}^{k}\frac{PR\_W(p_i)}{C(p_i)} \quad (2)$$

where $CC(p)$ is the number of citations pointing to author p, $\sum_{j=1}^{N}CC(p_j)$ is the citation counts of all nodes in the network, and (1-d) is the coefficient to retain the sum of PageRank as one. Let $\overline{\overline{M}}$ be the PageRank matrix; let $\overline{M}$ be the stochastic matrix for the adjacency matrix with the rows and columns corresponding to the directed graph of the coauthorship network; let *e* be the n-vector whose elements are all $e_i = 1$ and *v* is an n-vector, or referred to as personalized vector (Haveliwala, Kamvar, & Jeh, 2003), whose elements are $\frac{CC(p)}{\sum_{j=1}^{N}CC(p_j)}$; and let *x(v)* be the personalized PageRank vector corresponding to the personalization vector *v*. Based on this, *x(v)* can be computed by solving $x = \overline{\overline{M}}x$ (Haveliwala et al., 2003), where $\overline{\overline{M}} = d\overline{M} + (1-d)ve^T$. Therefore, *x* can be calculated as:

$$x = (1-d)(I - d\overline{M})^{-1}v \quad (3)$$

By letting $N = (1-d)(I - d\overline{M})^{-1}$, then $x = Nv$. According to Haveliwala et al. (2003), *N* comprises a complete basis for personalized PageRank vectors, since any personalized PageRank vector can be expressed as a convex combination of the columns of *N*. For any *v*, the corresponding personalized PageRank vector is given by *Nv*.

*PR_W* provides an integrated algorithm to combine citation and the topology of the network in a simple and efficient way. In one extreme case, when damping factor *d* equals zero, each node would have its relative citation score: $\frac{CC(p)}{\sum_{j=1}^{N} CC(p_j)}$ which equals the normalized citation counts. Another extreme case is when *d* equals to one; however, if *d* is too close to 1, then the PageRank may become unstable and the convergence rate slows (Boldi et al., 2005). Based on this, we choose damping factor of 0.85, 0.75, 0.65, 0.55, 0.45, 0.35, 0.25, and 0.15 respectively, and rank authors under these circumstances. In order to differentiate the original PageRank (formula (1)) and the proposed weighted PageRank (formula (2)), we use PR to refer to the original PageRank and PR_W for the weighted PageRank in the following paragraphs, and the damping factor used is enclosed in parenthesis. For example, e.g. PR_W(0.85) is weighted PageRank under damping factor 0.85.

## 4 Results and analysis

### *4.1 Correlation between PageRank and citation rank*

Before calculating the correlation between PR and citation rank, we present the concept of component. In social network analysis, connected graphs are referred to as components. A component of a graph is a subset with the characteristic that there is a path between one node and any other nodes in the same subset (Nooy, Mrvar, & Batagelj, 2005). A coauthorship network generally consists of many disconnected components, and we usually focus on the largest components, since metrics like closeness centrality, mean distance, and clustering coefficient can only be applied to a single, connected component. As for PR, Liu et al. (2005) applied it to a whole network. But in this study, we only apply PR to the largest component which contains 1,034 authors, since smaller components would usually yield unproportionate PR values and would result in too much noise for the ranking results.

In the following part of this section, we study the distribution of citation counts and PR scores and the correlation between them. Redner (1998) studies two sets of citation distribution data: (1) 783,339 papers cataloged by ISI in 1981 and (2) 24,296 papers published in Physical Review D between 1975 and 1994. For both of the data sets, he found that the probability a paper had been cited *k* times followed a power law $p(k) \sim k^{-\lambda}$ with exponent $\lambda = 3$. It indicates that the in-degree distribution of the citation network follows a power law. A similar study by Vázquez (2001) found the out-degree distribution had an exponential tail. In this study, we find citation counts and PR values also follow power-law distribution, with R equals 0.8504 (citation) and 0.9083 (PR) respectively (shown in Figure 1). Hence, PR can also be a substitutive indicator, which can also reflect the scale-free characteristic in scientific collaboration networks.

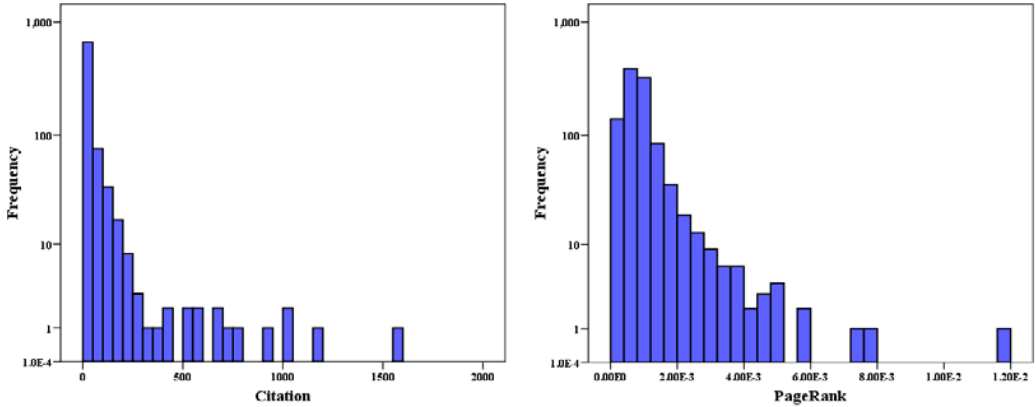

Figure 1. Distribution of citation and PR(0.85)

Currently, studies on PR for bibliometrics calculate overall correlation between citation and PR. For example, Ma et al. (2008) found the correlation between paper citation counts and paper PR values is around 0.9; Bollen et al. (2006) found their weighted PR values from journal citation network and Journal Citation Report impact factors are correlated, with Spearman's correlation 0.61. Fiala et al. (2008) also found citation, PR, HITS, and another seven self-defined weighted PR are highly correlated with each other. In this study, we take a different perspective by stratifying the ranking levels and calculating correlations within each level, as shown in Table 1 (rankings for each level are based on citations).

Table 1. Spearman's correlation between PageRank and citation for each ranking level*

| Ranking Levels (obverse) | | | | | | | |
| --- | --- | --- | --- | --- | --- | --- | --- |
| | 1~30 | 1~50 | 1~100 | 1~200 | 1~300 | 1~500 | 1~1034 |
| Spearman's | 0.3672 | 0.4166 | 0.5278 | 0.5022 | 0.4490 | 0.4747 | 0.4673 |
| P value | 0.0042 | 0.0013 | 0.0000 | 0.0000 | 0.0000 | 0.0000 | 0.0000 |
| Ranking Levels (reverse) | | | | | | | |
| | 1034~500 | 1034~300 | 1034~200 | 1034~100 | 1034~50 | 1034~30 | 1034~1 |
| Spearman's | 0.0540 | 0.1619 | 0.2592 | 0.3269 | 0.3869 | 0.4119 | 0.4673 |
| P value | 0.1683 | 0.0028 | 0.0001 | 0.0000 | 0.0000 | 0.0000 | 0.0000 |

*$d$=0.85

For obverse ranking levels, the overall correlation is 0.4673, and the highest correlation occurs in the top 100 level. For reverse ranking levels, correlations decrease from 0.4673 to 0.0540 as levels move to the back part of the ranking. Low correlation in the tail part indicates that PR values for authors in these levels are inconsistent, in that these values are small in their scale and very susceptible to fluctuations: a little higher or lower for citation counts would result in a quite significant change for PR. We therefore argue that for the PR algorithm, only the top 10%-20% of overall authors in the coauthorship network can produce useful data.

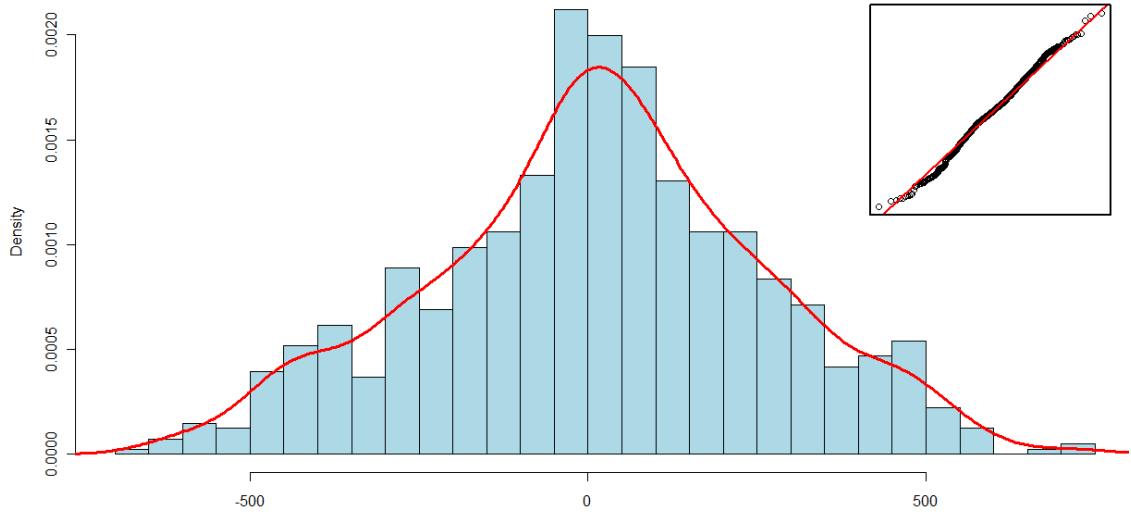

Figure 2. Rank changes between citation and PR(0.85) with quantile-quantile diagram

Figure 2 presents the rank changes between citation and PR. The majority variances are located near zero, and only a portion has diverse rank status. The quantile-quantile diagram also suggests that the rank variance follows a normal distribution: there are fewer records as the variance increases. The overall correlation between citation and PR is 0.4673, which is lower when compared to journal citation networks r=0.61 (Bollen et al., 2006) and paper citation networks r=0.9 (Ma et al., 2008). This is resulted from the type of networks under study. Links in coauthorship network are coauthor relations, whereas links for journal citation network and paper citation network are citation relations. These networks are thus more pertinent to citations, and their having higher correlation with citation counts comes as no surprise.

Figure 3 shows the scatter plot of citation counts and PR values. In this log-log graph, more nodes are distributed around the diagonal line, indicating that PR values and citation counts are correlated. Nevertheless, for lower values, each citation count covers a wide range of PR values, and consequently their PR values and citation counts do not have correlation with each other, and therefore they are distributed horizontally rather than diagonally.

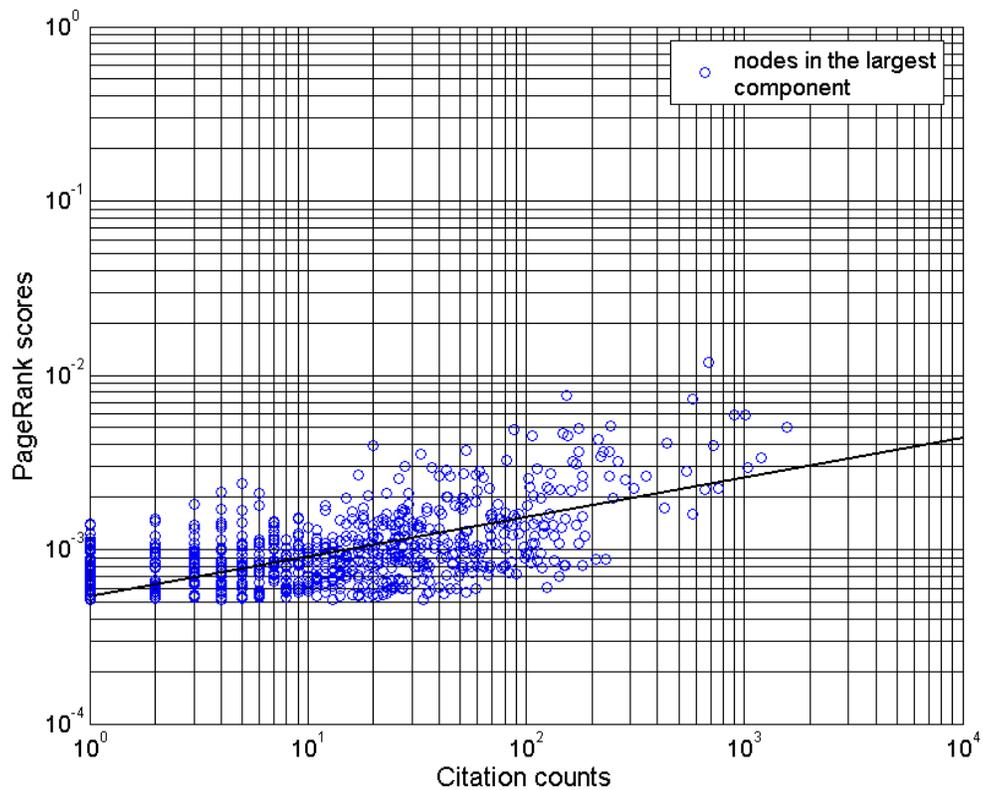

Figure 3. PR(0.85) and citation counts

## *4.2 PageRank with different damping factors*

In Figure 4, we present the Spearman's rank correlation and scatter plot for different damping factors. We find that PR scores for different damping factors are significantly correlated, with correlation coefficients for all pairs above 0.95. The correlation coefficients of PR are decreasing with the increment of intervals for damping factors: the correlations for adjacent damping factors are all greater than 0.99, and the lowest correlation exists between the most remote pair 0.85 and 0.15 ($r_s$=0.95). We also find that when damping factor changes, the top and tail part still converge to the distribution lines, but the middle part has some discrepancies (the shuttle shape). This result is consistent with Ding et al.'s finding (2009) who found that damping factors have limited effects on a co-citation network, which differs from Pretto's finding (2002) that when *d* changes, the lower section of the ranking varies dramatically.

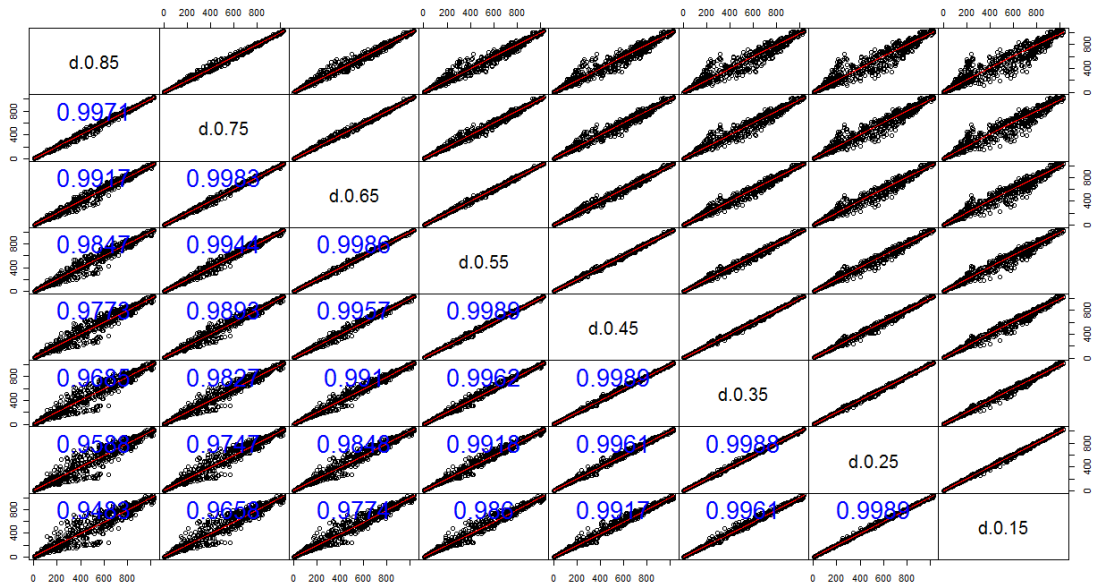

Figure 4. Distribution and correlation of PR values under different damping factors

*4.3 Weighted PageRank*

The PR itself (equation (1)) is a weighted algorithm; it recursively redistributes link weight until all values converge. For coauthorship networks, however, the PR may not be accurate enough for evaluating author impact. A coauthorship network only reflects an author's scientific collaboration status, and therefore single authored or authors who do not actively participant in scientific collaboration would have disadvantages. In this study, we extend PR by incorporating citation into PR algorithm, as shown in equation (2). Our version of PR_W integrates an author's community impact (via coauthorship) and academic impact (via citation), which may be a more reliable measure for impact analysis. We also calculate PR_W values under different damping factors, and list the top 20 authors for each situation.

Table 2. Top 20 authors for weighted PageRank under different damping factors

|    | PR_W(0.85)    | PR_W(0.75)    | PR_W(0.65)    | PR_W(0.55)    | PR_W(0.45)    | PR_W(0.35)    | PR_W(0.25)    | PR_W(0.15)    |
|----|---------------|---------------|---------------|---------------|---------------|---------------|---------------|---------------|
| 1  | Rousseau, R   | Rousseau, R   | Glanzel, W    | Glanzel, W    | Glanzel, W    | Glanzel, W    | Glanzel, W    | Glanzel, W    |
| 2  | Glanzel, W    | Glanzel, W    | Rousseau, R   | Schubert, A   | Schubert, A   | Schubert, A   | Schubert, A   | Schubert, A   |
| 3  | Moed, HF      | Moed, HF      | Moed, HF      | Rousseau, R   | Moed, HF      | Moed, HF      | Moed, HF      | Braun, T      |
| 4  | Thelwall, M   | Thelwall, M   | Schubert, A   | Moed, HF      | Rousseau, R   | Braun, T      | Braun, T      | Moed, HF      |
| 5  | Leydesdorff, L| Schubert, A   | Thelwall, M   | Thelwall, M   | Braun, T      | Thelwall, M   | Thelwall, M   | Thelwall, M   |
| 6  | Schubert, A   | Leydesdorff, L| Braun, T      | Braun, T      | Thelwall, M   | Rousseau, R   | Rousseau, R   | Rousseau, R   |
| 7  | Oppenheim, C  | Braun, T      | Leydesdorff, L| Leydesdorff, L| McCain, KW    | McCain, KW    | Egghe, L      | Egghe, L      |
| 8  | Braun, T      | McCain, KW    | McCain, KW    | McCain, KW    | Leydesdorff, L| Egghe, L      | McCain, KW    | McCain, KW    |
| 9  | McCain, KW    | Oppenheim, C  | Egghe, L      | Egghe, L      | Egghe, L      | Leydesdorff, L| Leydesdorff, L| White, HD     |
| 10 | Courtial, JP  | Egghe, L      | Oppenheim, C  | White, HD     | White, HD     | White, HD     | White, HD     | Leydesdorff, L|
| 11 | Cronin, B     | Cronin, B     | White, HD     | van Raan, AFJ | van Raan, AFJ | van Raan, AFJ | Small, H      | Small, H      |
| 12 | Lewison, G    | White, HD     | Van Raan, AFJ | Ingwersen, P  | Ingwersen, P  | Small, H      | van Raan, AFJ | van Raan, AFJ |
| 13 | Egghe, L      | van Raan, AFJ | Ingwersen, P  | Small, H      | Small, H      | Ingwersen, P  | Ingwersen, P  | Ingwersen, P  |
| 14 | Kretschmer, H | Ingwersen, P  | Cronin, B     | Cronin, B     | Cronin, B     | Cronin, B     | Cronin, B     | Cronin, B     |
| 15 | Kostoff, RN | Courtial, JP | Small, H | Oppenheim, C | Oppenheim, C | Persson, O | Persson, O | Persson, O |
| 16 | Gupta, BM | Lewison, G | Courtial, JP | Courtial, JP | Persson, O | Oppenheim, C | Harter, SP | Harter, SP |
| 17 | Wilson, CS | Small, H | Lewison, G | Persson, O | Courtial, JP | Harter, SP | Nederhof, AJ | Nederhof, AJ |
| 18 | van Raan, AFJ | Kostoff, RN | Kostoff, RN | Kostoff, RN | Harter, SP | Courtial, JP | Courtial, JP | Vaughan, L |
| 19 | Ingwersen, P | Kretschmer, H | Persson, O | Lewison, G | Kostoff, RN | Nederhof, AJ | Vaughan, L | Courtial, JP |
| 20 | White, HD | Wilson, CS | Chen, CM | Harter, SP | Vaughan, L | Vaughan, L | Oppenheim, C | Chen, CM |

For *d*=0.85, the topology of the coauthorship network plays a major role, while for *d*=0.15, citation counts are more influential. For example, Rousseau, R and Kretschmer, H have collaborated with authors from varied regions, and hence they ranked higher for *d*=0.85; for *d*=0.15 where citation takes a dominant role, authors whose works are highly cited would rank higher, such as White, HD and Small, H. Twenty-six authors are recorded in Table 2, of which 14 (in alphabetical order: Braun, T, Courtial, JP, Cronin, B, Egghe, L, Glanzel, W, Ingwersen, P, Leydesdorff, L, McCain, KW, Moed, HF, Rousseau, R, Schubert, A, Thelwall, M, van Raan, AFJ, White, HD) are listed in the top 20 for all damping factors. These authors have the highest academic impact and community impact, and are mainstays in the informetrics research community. This result also suggests that this PR_W is quite reliable, where the majorities remain stable for varied damping factors. Accordingly, this PR_W provides an integrated algorithm to combine citation and the topology of the coauthorship network in an integrated measure.

*4.4 Evaluation*

We collected the PC membership data for 12 ISSI conferences and showed the number of times those authors were PC members in the column "PC member" (attendance data for the first two conferences). We chose three PR_Ws: PR_W(0.15) where citation plays a major role, PR_W(0.55) where the coauthorship network topology and author's citation counts have nearly the same impact towards the PR_W score, and PR_W(0.85) where the coauthorship network topology plays a major role.

Table 3. Comparison of different metrics

| Author | Rank | | | | | | Citation counts* | $h$-index | PC member |
|---|---|---|---|---|---|---|---|---|---|
| | PR_W(0.55) | PR(0.55) | PR_W(0.15) | PR(0.15) | PR_W(0.85) | PR(0.85) | | | |
| **Glanzel, W** | 1 | 9 | 1 | 11 | 2 | 7 | 1571 | 26 | 12 |
| **Schubert, A** | 2 | 26 | 2 | 26 | 6 | 23 | 1191 | 20 | 1 |
| **Rousseau, R** | 3 | 1 | 6 | 1 | 1 | 1 | 687 | 16 | 12 |
| **Moed, HF** | 4 | 5 | 4 | 5 | 3 | 5 | 1014 | 19 | 10 |
| Thelwall, M | 5 | 4 | 5 | 4 | 4 | 4 | 904 | 17 | 4 |
| **Braun, T** | 6 | 34 | 3 | 39 | 8 | 29 | 1040 | 19 | 9 |
| **Leydesdorff, L** | 7 | 3 | 10 | 3 | 5 | 3 | 583 | 15 | 5 |
| **McCain, KW** | 8 | 12 | 8 | 10 | 9 | 16 | 722 | 13 | 7 |
| **Egghe, L** | 9 | 50 | 7 | 46 | 13 | 55 | 761 | 16 | 10 |
| **White, HD** | 10 | 68 | 9 | 86 | 20 | 56 | 658 | 11 | 3 |
| **van Raan, AFJ** | 11 | 27 | 12 | 25 | 18 | 34 | 543 | 19 | 9 |
| **Ingwersen, P** | 12 | 32 | 13 | 31 | 19 | 35 | 541 | 11 | 8 |
| **Small, H** | 13 | 93 | 11 | 90 | 24 | 100 | 579 | 11 | 7 |

| | | | | | | | | |
|---|---|---|---|---|---|---|---|---|
| Cronin, B | 14 | 15 | 14 | 12 | 11 | 14 | 441 | 12 | 7 |
| Oppenheim, C | 15 | 2 | 22 | 2 | 7 | 2 | 154 | 8 | 1 |
| Courtial, JP | 16 | 7 | 19 | 9 | 10 | 6 | 243 | 8 | 2 |
| Persson, O | 17 | 38 | 15 | 38 | 23 | 42 | 355 | 11 | 3 |
| Kostoff, RN | 18 | 10 | 21 | 8 | 15 | 13 | 213 | 9 | 1 |
| Lewison, G | 19 | 6 | 23 | 6 | 12 | 8 | 174 | 8 | 5 |
| Harter, SP | 20 | 39 | 16 | 33 | 26 | 53 | 311 | 6 | 1 |

*Citation and *h*-index are subject to the data of this study

When a new metric is proposed, it is natural to try to evaluate the rankings calculated by it. A most straightforward solution would be to compare the generated rankings with an unassailable official ranking; unfortunately, this may not be applicable in most circumstances. Since citation is so prevalent, it has become a common practice to compare new rankings with citation rankings. But for this study, citation has been incorporated into the PR_W; therefore, it is not statistically justifiable to compare with citation in this instance. A possible alternative is to compare the new ranking indirectly through awards lists, as practiced by Sidiropoulos and Manolopoulos (2006) and Fiala et al. (2008). Quite logically, award winners are expected to rank higher than other authors (Sidiropoulos & Manolopoulos, 2006). For this study, we compare the PR_W with the Derek de Solla Price Award (http://www.issi-society.info/price.html) rewarded by ISSI. In Table 3, authors are displayed in bold font if they are Price Award winners. Of the top 13 authors based on the PR_W(0.55), 12 are Price Award winners, which share the same results if ranked by citation counts. This is reasonable since prestige, popularity, awards, and recognition still rely mostly on the number of an author's citations (Fiala et al., 2008). In addition, PR_W(0.15) includes 12 Price Award winners in the top 13 list; PR_W(0.85) includes 12 winners in the top 24 list; however, PR(0.15) includes 12 winners only in the top 90 list; PR(0.55) includes 12 winners only in the top 93 list; PR(0.85) includes 12 winners only in the top 100 list. PR_W thus has a better match with Price Award winners than the PR, *h*-index, and PC member.

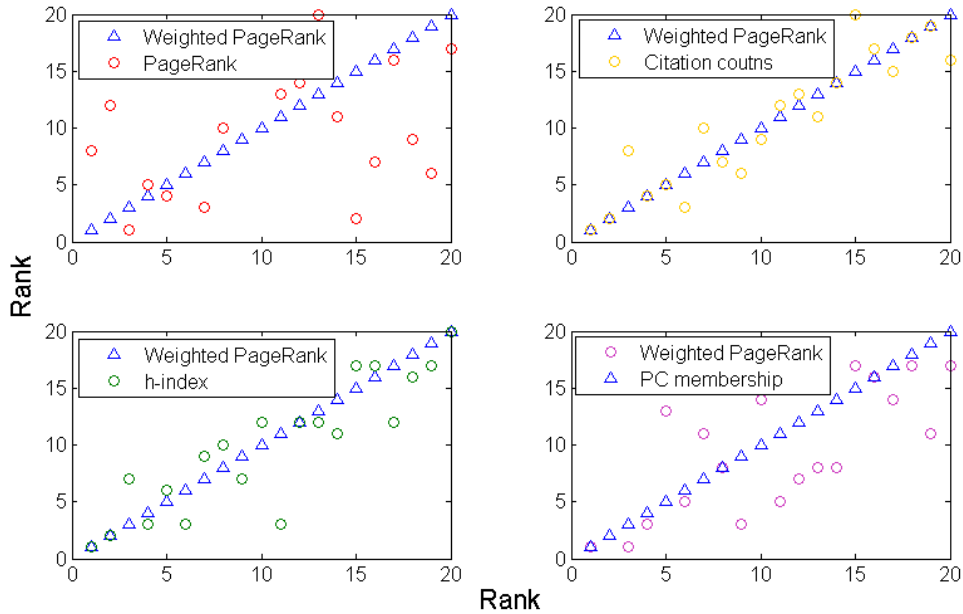

Figure 5. Scatter plot of different metrics (*d*=0.55)

Figure 5 shows the scatter plot of *h*-index, PR(0.55), Citation, and PC members with PR_W(0.55) for top 20 authors (see Table 3). Citation and *h*-index match with PR_W(0.55) more precisely, while PR(0.55) and PC member have some discrepancies. This is not surprising since PR, like PC membership, is more related to community impact, whereas *h*-index and citation concentrate on academic impact. The PR_W combines these two impacts by integrating citation (academic impact) and coauthorship network topology (community impact), shown as the linear lines in the center.

## 5 Conclusion

The current study applies PR to coauthorship network analysis. The PR algorithm provides a meaningful extension to the traditionally used citation counts for authors.

Through the correlation analysis between PR and citation for each author, we find that PR and citation are correlated and that PR, to a certain degree, also measures an author's academic impact. But they also differ when comparing to the significant correlation coefficients of paper citation networks and journal citation networks. This discrepancy is resulted from varied network formations: the former is based on coauthor ties, whereas the last two are based on citation relations that are more pertinent to citation counting. We also compute correlation coefficients for authors at different citation ranking levels, and find that PR values and citation counts have fewer correlations for the lower levels, which indicates that the PR algorithm may only yield useful values for the top 10% to 20% of authors. This result can be interpreted by the power-law distribution of PR values and citations, where only a few authors regularly participate in scientific collaboration.

The correlations between PR values and citation counts under different damping factors are stable, and the correlations for PR within different damping factors are significantly correlated with correlation coefficients all above 0.95. Therefore, for coauthorship networks, damping factors do not have much influence for PR values.

The PR_W combines citation and coauthorship network topology in a very effective way. Compared to other related PR algorithms, it focuses on the random surfing aspect and develops it into citation ratios. Accordingly, it integrates an author's community impact and academic impact, providing an alternative measure for identifying author impact in the informetrics community. Through comparison with Price Award winners, this PR_W accurately identifies these winners in the top of the ranking list and outperforms *h*-index and PR. Tested under different damping factors, there is little global reordering of authors. Therefore, this weighted algorithm is a reliable approach for author impact evaluation.

## Acknowledgement

The authors are indebted to Ronald Rousseau and Liming Liang who provided valuable conference data to this study.